\documentclass[aps,prd,showpacs,nofootinbib,floats,floatfix,preprintnumbers,groupedaddress,twocolumn]{revtex4}

\usepackage{bm}
\usepackage{latexsym}
\usepackage{dcolumn}
\usepackage{amsmath,amsfonts,amssymb}
\usepackage{graphicx,epsfig}
\usepackage{amsmath}
\usepackage{fancyhdr}
\usepackage{hyperref}
\usepackage{graphicx,epstopdf}
\hypersetup{
	colorlinks   = true, 
	urlcolor     = blue, 
	linkcolor    = blue, 
	citecolor   = red 
}

\def\r{\ref}
\def\no{\nonumber}
\def\a{\alpha}
\def\b{\beta}
\def\g{\gamma}
\def\d{\delta}

\def\r{\ref}
\def\p{\partial}
\def\f{\frac}

\begin{document}
\title{Phase transition and critical phenomena of black holes: A general approach}
\author{Abhijit Mandal$^{a}$\footnote {\color{blue} abhijitmandal.math@gmail.com}}
\author{Saurav Samanta$^{b}$\footnote {\color{blue} srvsmnt@gmail.com}}
\author{Bibhas Ranjan Majhi$^{c}$\footnote {\color{blue} bibhas.majhi@iitg.ernet.in}}

\affiliation{$^a$Department of Mathematics, Jadavpur University, Kolkata 700032, West Bengal, India\\
$^b$Department of Physics, Narasinha Dutt College, 129, Belilious Road, Howrah 711101, India\\
$^c$Department of Physics, Indian Institute of Technology Guwahati, Guwahati 781039, Assam, India
}

\date{\today}

\begin{abstract}
We present a general framework to study the phase transition of a black hole. Assuming that there is a phase transition, it is shown that without invoking any specific black hole, the critical exponents and the scaling powers can be obtained. We find that the values are exactly same which were calculated by taking explicit forms of different black hole spacetimes. The reason for this universality is also explained. The implication of the analysis is -- {\it one does not need to investigate this problem case by case}, what the people are doing right now. We also observe that these, except two such quantities, are independent of the details of the spacetime dimensions. 
\end{abstract}

\maketitle

    {\bf{Introduction}}:
    Works of Bekenstein \cite{Bekenstein:1973ur} and Hawking \cite{Hawking:1974sw} revealed that black holes behave like ordinary thermodynamical objects (see also \cite{Bardeen:1973gs}). One can associate entropy and temperature on the horizon and the laws of black hole mechanics are quite analogous to those of our usual thermodynamics. Though we still lack a microscopic explanation of these thermodynamic entities (like entropy, temperature etc.), we can use semi-classical treatment of gravitational theories. Research of past few years reveals that, {\it such a thermodynamical similarity of black holes is quite an universal feature -- irrespective of the gravitational theories}. So, even in absence of full quantum description, at this point we are still confident about the basic equations (although in some cases one has to carefully check it, like in dynamical spacetimes). And we use these equations to probe deep inside black hole mechanics, provided we identify the macroscopic parameters properly.
    
    One of the main research area which people emphasize a lot is phase transition and its consequences. This has been started by the seminal work of Davis \cite{Davies:1978mf}. He claimed a certain type of phase transition by observing the discontinuity of specific heat. Later on it has been argued that existing information is insufficient to treat it as a phase transition \cite{Kaburaki:1993ah}. Subsequently, a phase transition was observed by Hawking and Page \cite{Hawking:1982dh} -- the Schwarzschild-AdS black hole exhibits first order phase transition at high temperature.  It must be mentioned that the asymptotic AdS nature is crucial here, because for asymptotically flat case the specific heat is negative and hence thermodynamically unstable. This type of phase transition is commonly known as {\it Hawking -- Page phase transition}. Consequently, other types of phase transition have also been proposed. Weinhold and Ruppeiner's \cite{Weinhold,Ruppeiner:1995zz} proposed a method by interpreting the geometric quantities (mainly the metric coefficients) as thermodynamic variables. A transition from non-extremal to extremal black hole was observed by finding the divergence of inverse specific heat (not the divergence of specific heat) \cite{Kaburaki,Cai:1996df}. An interesting observation for the last type of phase transition is that the critical exponents are same for few different black holes \cite{Cai:1996df}.
    
    Later on it turned out that, phase transition of black holes with two or more macroscopic parameters can be studied in the usual framework of thermodynamics \cite{Banerjee1}--\cite{Mo}. In this case one looks for the divergence of specific and inverse isothermal compressibility. Interestingly, in this case one can study the critical phenomenon and consequently one can also find the critical exponents. Moreover, it turned out that the exponents obey the usual thermodynamical scaling laws. We shall call it Type I. Recently, another way of looking at the problem has been put forwarded \cite{Kubiznak:2012wp}--\cite{Zhao:2013oza}. This is mainly restricted to AdS black holes. In this case the cosmological parameter is treated as pressure and the whole system is mapped with the Van der Waals gas system. These will be named as Type II. Consequently, a great attention has been paid over the last few years on the critical behaviour of different black hole solutions in the framework of usual classical gravity or semi-classical gravity (See \cite{Banerjee1, Banerjee2, Majhi, Ma, Mo} for Type I and \cite{Kubiznak:2012wp} --\cite{Zhao:2013oza} for Type II). For more on phase transitions of different black holes, see \cite{more}.
    
       In the last two approaches, the critical exponents have been successfully calculated. People then started studying different black hole spacetimes. {\it Most noticeable feature of all these studies is that the value of critical exponents are same, irrespective of particular metric}. Of course, they do not match with those evaluated in other approach; {\it i.e.} the exact values of the critical exponents in Type I is independent of specific choice of black hole metric and same can be said for Type II. But there is no match between Type I and Type II. Such a typical universality within any one of the types, seems to be hiding in the methodologies itself.  It is like the fact that the universality of thermodynamical structure of spacetimes can be explained within a particular gravitational theory in a general approach, like Noether prescription by Wald \cite{wald,Majhi:2011ws} under the assumption that there exists a timelike bifurcated Killing horizon. Therefore it would be interesting to know the reason behind this characteristic. {\it Moreover, one would be very much eager to see if there exists a general way to find the critical exponents, without considering any specific metric, by assuming some underlying common features which are all contained in the black hole metric}. The necessity of such discussion is two fold. ($i$) First, we shall be able to understand why such an universality exists in phase transition. ($ii$) Second, we do not need to study case by case which people are doing right now.  

     As we said just now our aim in this paper is two fold. So the initial task is to identify the underlying common inputs in any one type (Type I or Type II) of phase transition. Based on that, assuming there is a phase transition, we shall see if a general analysis can be given to find the critical exponents without considering any specific black hole metric. In this paper we shall concentrate on Type I; the discussion on Type II will be elsewhere. It will be observed that such a desired approach is indeed possible. The critical exponents  exactly matches with the earlier findings by studying case by case {\footnote{The same values of critical exponents were also obtained in an old work \cite{Cai}.}}. In this way we shall also notice that, the calculation is very simple compared to that for any specific metric as we do not need to consider the complicated forms of thermodynamical entities.

\vskip 2mm
\noindent
{\bf{Critical Phenomena in a general frame work}}:
   The first law of thermodynamics is 
 $dE=TdS-\sum_iY_idX_i$
  where $E$ is internal energy, $T$ is temperature and $S$ is entropy of the system. Other two sets of variables $Y_i$ and $X_i$ are generalized force and appropriate generalized displacement. For a gaseous system $Y$ is pressure and $X$ is volume. For a magnetic system $Y$ is magnetic intensity and $X$ is magnetization. In the case of black hole, electric potential $\Phi$ and angular velocity $\Omega$ are generalized forces. Whereas, electric charge $Q$ and angular momentum $J$ are the generalized displacements for $\Phi$ and $\Omega$ respectively. In this article, for simplicity, we shall consider systems with either ($Y=\Phi,~X=Q$) or ($Y=\Omega,~X=J$).
For a large number of black holes, event horizon $r_+$ and $X$ can be viewed as only independent variables. Other thermodynamical variables can be conveniently written as their functions. For example,
$S=S(r_+)$, $Y=Y(X,r_+)$, $T=T(X,r_+)$.
  
 In order to study phase transition, one looks at the values of specific heat ($C_X$) and the inverse of isothermal compressibility $(K_T^{-1})$. These are defined as
$C_X =  T\left( {\p S}/{\p T} \right)_X; \,\,\,\  K_T^{-1}=X\left( {\p Y}/{\p X} \right)_T$.
If for some particular values of $r_+(=r_{+c})$, $C_X$ and $K_T^{-1}$ diverge that indicates a phase transition (first or higher order) of the corresponding black hole. That particular point ($r_{+c}$) is called critical point. 
   
   In general, it is not easy to express $S$ (or $Y$) in terms of $X$ and $T$. So one uses the chain rule of partial differentiation to express $C_X$ in the following equivalent way,
\begin{equation}
C_X=T\left( \frac{\p S}{\p r_+} \right)_X\left( \frac{\p T}{\p r_+} \right)_X^{-1}~.
\label{CX}
\end{equation}
Similarly we have $\left( {\p Y}/{\p X} \right)_T=\left( {\p Y}/{\p X} \right)_{r_+}+\left( {\p Y}/{\p r_+} \right)_X\left( {\p r_+}/{\p X} \right)_T$. 
At this point we recall, for any function $f(r_+,X,T)=0$ there exists an important rule of partial differentiation: $\left( {\p r_+}/{\p X} \right)_T\left( {\p X}/{\p T} \right)_{r_+}\left( {\p T}/{\p r_+} \right)_X=-1$.
Using these we express $K_T^{-1}$ as,
\begin{equation}
K_T^{-1}=X\left( \frac{\p Y}{\p X} \right)_{r_+} - X\left( \frac{\p Y}{\p r_+} \right)_X\left( \frac{\p T}{\p X} \right)_{r_+} \left( \frac{\p T}{\p r_+} \right)_X^{-1}
\end{equation}
Clearly, when $\left( {\p T}/{\p r_+} \right)_X=0$ both $C_X$ and $K_T^{-1}$ diverge (assuming no other singular term in the expression). This is a typical signature of phase transition. 

	Now the question is: can we find the critical exponents for phase transition without any explicit information about a black hole?  Till now people have looked into this problem case by case. Here we shall show that without invoking any explicit black hole spacetime, a general analysis is indeed possible. Let us follow the line of usual approach. Near the critical point, where phase transition occurs, we denote the values of the macroscopic parameters as
$T\rightarrow T_c,~~r_+\rightarrow r_{+c}~~\mbox{and}~~X\rightarrow X_c$,
where suffix `$c$' denotes the value of the physical quantity at the critical point.
Then we re-express the physical quantities near the critical points as
\begin{eqnarray}
&& r_+=r_{+c}(1+\Delta)~, \,\,\
T(r_+)=T(r_{+c})(1+\epsilon)~,
\nonumber
\\
&& X(r_+)=X(r_{+c})(1+\Pi)~,
\label{rTX}
\end{eqnarray}
where $\Delta<<1$, $\epsilon<<1$ and $\Pi<<1$. Next idea is to expand the different quantities around this critical point. 

     Let us start with the horizon temperature ($T$). Perform the Taylor series expansion of $T$ for a fixed value of $X$ in the neighbourhood of $r_{+c}$. 
Now, since the signature of phase transition is zero value of $(\partial T/\partial r_+)_X$ at the critical point, keeping upto second order term and using (\ref{rTX}) we get
$\epsilon T_c=\f{1}{2}\left[\left( {\p^2 T}/{\p r_+^2} \right)_X\right]_{r_+=r_{+c}}r_{+c}^2\Delta^2$
where $T_c\equiv T(r_{+c})$.
Therefore the order of $\Delta$ in terms of $\epsilon$ goes as
\begin{equation}
\label{Del_epsilon}
\Delta\sim\sqrt{\epsilon T_c}~.
\end{equation}
Since for our main purpose the derivative term and $r_{+c}$ are not needed, in the above those parts are not mentioned explicitly.
Similarly, the Taylor series expansion of $X$ for a fixed temperature ($T$) around the critical point, upto second order term, is
\begin{eqnarray}
X&=&X(r_{+c})+\left[\left( \frac{\p X}{\p r_+} \right)_T\right]_{r_+=r_{+c}}(r_+-r_{+c})\nonumber
\\
&+&\f{1}{2}\left[\left( \frac{\p^2 X}{\p r_+^2} \right)_T\right]_{r_+=r_{+c}}(r_+-r_{+c})^2~.\label{Q_taylor}
\end{eqnarray}
The second term can be expressed as
$\left( {\p X}/{\p r_+} \right)_T=-\left( {\p T}/{\p r_+} \right)_X\left({\p X}/{\p T} \right)_{r_+}$.
Therefore at the critical point  this will vanish. Then using (\ref{rTX}) one obtains
$X(r_{+c})\Pi=\f{1}{2}\left[\left( {\p^2 X}/{\p r^2} \right)_T\right]_{r_+=r_{+c}}r_{+c}^2\Delta^2$,
which implies
\begin{equation}
\label{Del_Pi}
\Delta\sim\sqrt{X_c\Pi}~.
\end{equation}

Now we are in a position to find the values of the critical exponents. As is well known, these exponents basically describe the singular behaviour of various thermodynamic quantities near the critical point. Their definitions are given by:
\begin{eqnarray}
&&C_X\sim |T-T_c|^{-\a},Y(r_+)-Y(r_{+c})\sim|T-T_c|^{\b},\no\\
&&C_X\sim |X-X_c|^{-\varphi},  Y(r_+)-Y(r_{+c})\sim|X-X_c|^{\f{1}{\d}}, \label{def}\\
&&K_T^{-1}\sim |T-T_c|^{-\g},S(r_+)-S(r_{+c})\sim|X-X_c|^{\psi}.\no
\end{eqnarray}

Using the Taylor series expansion of $\left( {\p T}/{\p r_+} \right)_X$ and then (\ref{rTX}) one finds
$\left( {\p T}/{\p r_+} \right)_X= \left[\left( {\p^2 T}/{\p r_+^2} \right)_X\right]_{r_+=r_{+c}}r_{+c}\Delta\sim \Delta$ (since at the critical point, $\left( {\p T}/{\p r_+} \right)_X=0$).
Next using (\ref{Del_epsilon}) we evaluate
\begin{equation}
\left( \frac{\p T}{\p r_+} \right)_X\sim\sqrt{\epsilon T_c}\sim|T-T_c|^{\f{1}{2}}~.
\label{new1}
\end{equation}
Note that near the critical point, this is the only divergent term in the expressions of $C_X$ and $K_T^{-1}$, which implies $C_X$ diverges as
\begin{equation}
C_X\sim|T-T_c|^{-\f{1}{2}}~.
\label{C_Q&epsilon}
\end{equation}
Therefore comparing with (\ref{def}) we identify one of the critical exponents as $\a = 1/2$. Similarly we find $K_T^{-1}\sim|T-T_c|^{-\f{1}{2}}$ and hence $\gamma=1/2$.
We already showed that $\Delta\sim\sqrt{\epsilon}$, $\Delta\sim\sqrt{\Pi}$ and therefore $\epsilon\sim\Pi$. Hence $C_X$ diverges as
\begin{equation}
C_X\sim\epsilon^{-1/2}\sim\Pi^{-1/2}\sim|X-X_c|^{-\f{1}{2}}~.
\label{K_T&Pi}
\end{equation}
Comparing this with the standard form (\ref{def}), we obtain $\varphi = 1/2$.

	To find the other exponents we need to concentrate on $Y$ and $S$. Like the earlier, we expand $Y$ in Taylor series around the critical point.
Neglecting the higher order terms and using (\ref{rTX}) we find
$Y(r_+)-Y(r_{+c})=\f{1}{2}\left[\left( {\p \Phi}/{\p r_+} \right)_X\right]_{r_+=r_{+c}}r_{+c}\Delta$.
Remembering the relation between $\Delta,~\epsilon$ and $\Pi$ (see discussion above Eq. (\ref{K_T&Pi})), one finds,
\begin{eqnarray}
Y(r_+)-Y(r_{+c})\sim\Delta\sim\sqrt{\epsilon}\sim\sqrt{\Pi}~.
\label{Phi_Pi_epsilon}
\end{eqnarray}
Therefore we have $Y(r_+)-Y(r_{+c})\sim |X-X_c|^{1/2}\sim |T-T_c|^{1/2}$ and so $\beta = 1/2$, $\delta = 2$.
Similarly, we consider the Taylor series expansion of entropy ($S$) for the fixed value of $X$ around the critical point. 
So near the critical point, neglecting the higher order terms  we obtain
\begin{equation}\label{S_Pi}
S(r_+)- S(r_{+c}) \sim (r_+ - r_{+c}) \sim\Delta\sim\sqrt{\Pi}\sim|X-X_c|^{\f{1}{2}}
\end{equation}
and hence $\psi = 1/2$. 
In the above table we give all the values of critical exponents.\\
\begin{table}[!htb]
\caption{\label{tab:table1}}
\begin{ruledtabular}
\begin{tabular}{cccccccc}
Critical exponents &$\a$ &$\b$ &$\g$
 &$\d$  &$\varphi$ &$\psi$\\
\hline
Values& 1/2 & 1/2 & 1/2 & 2
& 1/2 & 1/2  \\
\end{tabular}
\end{ruledtabular}
\end{table}\\
It can be checked that, the above mentioned values of critical exponents were obtained by taking explicit forms of black hole metric \cite{Banerjee1}--\cite{Mo}. Here we obtained them by a general analysis without considering any specific black hole spacetime. Also it is very interesting to note that, the same values of exponents were also obtained for a relativistic Bose gas, confined in a cubic box, when one considers the first order phase transition below a critical pressure (See \cite{Park:2010fg} for details). 

	Now we discuss about the thermodynamic scaling law \cite{Stanley1, Stanley2} for our present work. We know that, in usual thermodynamic systems critical exponents satisfy some relations among themselves, which are known as thermodynamic scaling laws. These relations are stated below:
\begin{eqnarray}
&&	\a + 2\b + \g = 2, \,\,\,\ \a + \b(\d +1) = 2,\nonumber \\	
&&  (2 - \a)(\d \psi -1)+ 1 = (1 - \a)\d, \no \\
&&  \g (\d + 1) = (2 - \a)(\d - 1),  \\
&&  \b \d = \b + \g, \quad \varphi + 2 \psi = 1/\d ,\no \\
&&  \d = \f{2 - \a + \g}{2 - \a - \g}, \,\,\ \varphi \b \d = \a, \,\,\ \psi \b \d = 1- \a~. \no 
\end{eqnarray}
These relations are not all independent of one another and in fact some of these equalities are predicted as inequalities under stability consideration by scaling hypothesis. 
It is interesting to note that, our obtained values of the critical exponents (see Table \ref{tab:table1}) satisfy all the above mentioned relations.
	
	Experimentally and theoretically it was found \cite{Lousto} that, the critical exponents do not depend on the details of a physical system. This observation is embodied in the universality hypothesis, which states that for a continuous phase transition the static critical exponents depend only on the following properties:
(1) the dimension of the system, 
(2) the internal symmetry dimension of the order parameter and
(3) the range of the interactions.
Here also we observed that, the obtained values do not care about the specific form of black hole metric. In fact, in the calculation we never considered any particular metric. This is because, near the critical point, results do not depend on the coefficients, appearing in the Taylor series expansions. So the universal nature of phase transition is also true for the black hole systems. Moreover, the way we have approached, it is clear that {\it the resultant values do not depend on the spacetime dimensions}.

	Now, we are going to calculate the additional scaling laws for the exponents $\nu$ and $\eta$ which are associated with the singularity of correlation length and correlation function. Again, near the critical point one can express Helmholtz free energy as $F = F_r + F_s$, where $F_r$ is the regular part of the Helmholtz free energy whose second order partial derivatives are well behaved near the critical point and $F_s$ is the singular part of the Helmholtz free energy which is responsible for the thermodynamic behaviour. In the neighbourhood of a critical point, the singular part of  Helmholtz free energy $F(T, X)$ is a generalised homogeneous function of its variables \cite{Stanley1}.
Since for a black hole, the Helmholtz free energy ($F$) is defined as
$F(T, X) = M - TS$.
Now, by generalised homogeneous function hypothesis \cite{Lousto, Wu, Stanley2}, we can write
\begin{equation}
\label{F_epsilon_Pi_initial}
F(\lambda^p \epsilon, \lambda^q \Pi) = \lambda F(\epsilon, \Pi),
\end{equation}
where, $p$ and $q$ are two parameters called scaling powers and $\lambda$ is arbitrary. To proceed, let us use the Taylor expansion of $F(T, X)$ in the neighbourhood of $r_{+c}$; {\it i.e.} around $T=T_c$ and $X=X_c$. 
Now since $C_X(= - T \left({\p^2 F}/{\p T^2}\right)_X)$ and $K_T^{-1}(=X \left({\p^2 F}/{\p X^2}\right)_T)$ diverge at the critical point, one can easily evaluate the singular part of $F$ as
\begin{equation}
F_s=-\frac{C_X}{2T_c}(T-T_c)^2+\frac{K_T^{-1}}{2X_c}(X-X_c)^2~.
\end{equation}
Using (\r{new1}) and (\r{S_Pi}) we express $(T-T_c)$ in terms of $\epsilon$ and $X-X_c$ in terms of $\Pi$ respectively. Then the above reduces to
\begin{equation}
\label{F_epsilon_Pi_final}
F_s=a(X, r_+)\epsilon^{\f{3}{2}}+b(X, r_+)\Pi^{\f{3}{2}}~,
\end{equation}
where the irrelevant terms $a$ and $b$ contains the coefficients in front of $(T-T_c)^2$ and so on.
Comparing (\ref{F_epsilon_Pi_initial}) and (\ref{F_epsilon_Pi_final}), we get $p=2/3$ and $q=2/3$. Interestingly, here both scaling powers $p$ and $q$ have same values.
	
	In standard thermodynamics, the relation between the scaling powers and critical exponents are expressed as
	\cite{Stanley1}:
	\begin{eqnarray}
	&\a = 2 - \frac{1}{p}, \quad \b = \frac{1 - q}{p}, \quad \d = \frac{q}{1-q},\no \\
	&\g = \frac{2q - 1}{p}, \quad \psi = \frac{1 -p}{q}, \quad \varphi = \frac{2p -1}{q}. \no 
	\end{eqnarray}
	One can easily verify that the above equations are fulfilled by our determined critical exponents and scaling powers. Two remaining critical exponents which we have not discussed so far are $\nu$ and $\eta$. They are associated with the divergence of correlation length ($\xi$) and correlation function ($ G(\vec{r})$) respectively: $G(\vec{r})\sim e^{-r/\xi} ~\textrm{at}~ T\ne T_c,~\xi\sim\epsilon^{-\nu}$ and $G(\vec{r})\sim r^{-d+2-\eta}, ~\textrm{at}~ T= T_c$
	where $d$ is spatial dimensions of the theory. Now we assume the additional scaling relations,
	\begin{equation}
	\g = \nu(2 - \eta), \quad 2 - \a = \nu d~.
	\end{equation}	 
Taking $d=3$, we get $\nu = 1/2$ and $\eta = 1$.
	
	Let us finish this with the following comments. It must be pointed out that all the critical exponents and the scaling powers have the desired universality -- {\it they are same for all black holes}. Moreover, these except $\nu$ and $\eta$, {\it are independent of the spacetime dimensions}. Therefore, like the usual thermodynamical systems, the behaviour of black holes, as far as phase transition is concerned,  is very much universal in nature. This is possible, as we already noticed in the calculation, because the partial derivatives evaluated at the critical point, which incorporates the actual information of the black hole, do not make any contribution to the desired exponents and scaling powers. 
	
\vskip 2mm
\noindent
{\bf{Conclusions:}}	
   Phase transition in thermodynamics is one of the most important features which explains the abrupt behaviour of some macroscopic properties from one  phase to another. Interestingly, black hole behaves exactly like a thermodynamical object and all the laws are similar to that of standard thermodynamics. So it is natural to look for the possibility of phase transition in a black hole. It may be pointed out that the critical exponents play an important role in such studies. People actually spent a lot of time to explore the same for black hole. It turned out that phase transition can be of several types; like from non-extremal to extremal, Van der Waals type, etc. And the investigation is done case by case {\it i.e.} by considering the explicit form of black holes metric. In all cases one common feature is that the critical exponents are same. Then it is natural to identify the reason for this universality.
   
   Here in this paper we have precisely addressed this issue for a particular approach, which we called earlier as Type I. It has been observed that the critical exponents, indeed, can be obtained by a general approach without making use of any explicit form of the metric. Only we assumed the existence of a phase transition. The reason for such universality, we found, is due to the coefficients which depends on the metric, do not contribute to these exponents. It is noteworthy that, an attempt has already been made to find the critical exponents in a general frame work \cite{Azreg-Ainou:2014twa}, but the analysis was based on a general form of a metric. Here we did not even need that. 
   
   The importance of this analysis lies in the fact that one does not now need to check case by case, which people usually do. More importantly, the underlying reason for such universality is well understood from the present analysis. Additionally, the universal scaling laws are also obtained. This must enlighten the area of phase transition in a black hole system. Recently, different values of the critical exponents (so called non-standard) have been observed, which do not satisfy all the scaling laws, for the black holes in higher curvature gravity theories in the approach of Type II \cite{Mann}. It would be interesting to look for the reasoning behind this and its consequences. Finally we hope that similar general arguments can also be performed for other types of phase transition. The work in this direction are now in progress.  
\vskip 4mm
\noindent
{\bf{Acknowledgments}}\\
The research of one of the authors (BRM) is supported by a START-UP RESEARCH GRANT (No. SG/PHY/P/BRM/01) from Indian Institute of Technology
Guwahati, India.

\end{document}